\documentstyle[prl,aps,multicol,epsf]{revtex}

\begin{document}
 \draft
 \title{ Directed electron transport through ballistic quantum dot under microwave radiation}
 \author{ Jing-qiao Zhang, Sergey Vitkalov }
 \address{Physics Department, City College of the City University of New York, New York 10031, USA}
 \author{Z.D. Kvon}
 \address{Institute of Semicondutor Physics, 630090 Novosibirsk, Russia}
 \author{J. C. Portal$^{1,2,3}$ }
 \address{ $^1$GHMFL, CNRS/MPI-FKF, BP 166, Grenoble Cedex 9, F-38042, France; 
$^2$INSA, Toulouse Cedex 4, 31077 France;
$^3$Institut Universitaire de France
}
 \author{A.Wieck }
 \address{ Angewandte Festkorperphysik D-44780, Bochum, Germany }
 \date{\today}
 \maketitle
 
 \begin{abstract}
 
 Rectification of microwave radiation by asymmetric, ballistic quantum dot is observed. The directed transport is studied at different frequency (1-40 GHz) temperatures (0.3K-6K)and magnetic field. Dramatic reduction of the rectification is found in magnetic fields at which the cyclotron (Larmor) radius of the electron orbits at Fermi level is smaller than the size of the quantum dot. It strongly suggests the ballistic nature of the observed nonlinear phenomena.  Both symmetric and anti-symmetric with respect to the magnetic field contributions to the directed transport are presented. We have found that the behavior of the symmetric part of the rectified voltage with the magnetic field is different significantly for microwaves with different frequencies. A ballistic model of the directed transport is proposed.

 \end{abstract}
 
 \pacs{PACS numbers: }
 
 \begin{multicols}{2}

 Nonlinear directed transport in mesoscopic quantum objects has been investigated both theoretically and experimentally in different regimes \cite{thouless,falko,bykov,spivak,brouwer}. Most of the recent experimental efforts has been focused on the nonlinear rectification properties of quantum dots, where the electron transport is governed mostly by the electron quantum interference \cite{vavilov,dicarlo}. In this paper we present results of an experimental study of the directed electron transport through a quantum dot under microwave radiation in a regime, at which the classical ballistic trajectories appear to provide dominant contributions to the rectification effect. The dot with a lateral size $d \approx 1 \mu$ is studied. The size is significantly shorter than the mean free path $l_p=8 \mu$ of the 2D electrons in a complementary bulk system. We have observed a $gigantic$ reduction of the directed transport through the dot induced by external magnetic field.

The sample is fabricated from high mobility GaAs/AlGaAs
heterostructures using electron beam lithography and consecutive plasma etching. A schematic view of the sample is presented in the insert to figure 1. The dot is restricted by boundaries of three insulating disks of diameter 5 $\mu$. Two narrow conducting channels (the left and right channels on the insert) with a width approximately 0.3 $\mu $ and one channel with width 0.2 $\mu$ (the top, vertical channel on the insert) are formed between the disks. These channels provide electrical connections between the dot and three electrically isolated macroscopic 2D electron systems. A semitransparent metal gate was placed on the top of the structure.  Application of a negative voltage to the gate changed resistance of the channels. The channel with 0.2 $\mu$ width was most sensitive to the gate voltage $V_g$.  The bulk density of the 2DEG $n=3.08\times 10^{11}$cm$^{2}$ and the 2D bulk mobility $\mu= 0.91\times 10^{6}$cm$^{2}/$Vs at $V_g=0V$ did not vary considerably with the gate voltage. The bulk mean free path $l_p=8 \mu$ was much longer than the lateral size of the dot $d \approx 1.2\mu$(the dark circle in the insert). 

Measurements were done in vacuum chamber of He-3 insert at temperature 0.3 -6K and magnetic field up to 2T. The sample was placed on the top of a cooper plate attached to a He-3 chamber.  An additional cooling was provided by ten non-superconducting cold electrical wires thermally connected to the chamber. The microwave radiation (1-40 GHz) was applied through a semi-rigid coax, which was terminated by a parallel two-wire line. The line was placed at 3-5 mm above the sample. The electric field of polarized microwaves was oriented along the short edge of the sample chip as shown in the insert. The longitudinal resistance of bulk 2D part and the dot conductance were determined by a standard 4-points method at frequency 10 Hz in the linear regime. The rectified voltage was measured using high impedance digital multimeters across the dot between contacts 2 and 3 shown in the insert. 

 A dependence of the quantum dot conductance on the gate voltage  and on the magnetic field is presented in figure 1 (plots (A) and (B) on the left panel). The AC current was applied between the source and the drain. The AC voltage was measured between contacts 2-3. The dot conductance $G_{sd,23}$ is increasing with an increase of the negative gate voltage (curves from bottom  toward top on plot (B)).\\
\vbox{
\vspace{0.2 in}
\hbox{
\hspace{-0.2in} 
\epsfxsize 3.3 in \epsfbox{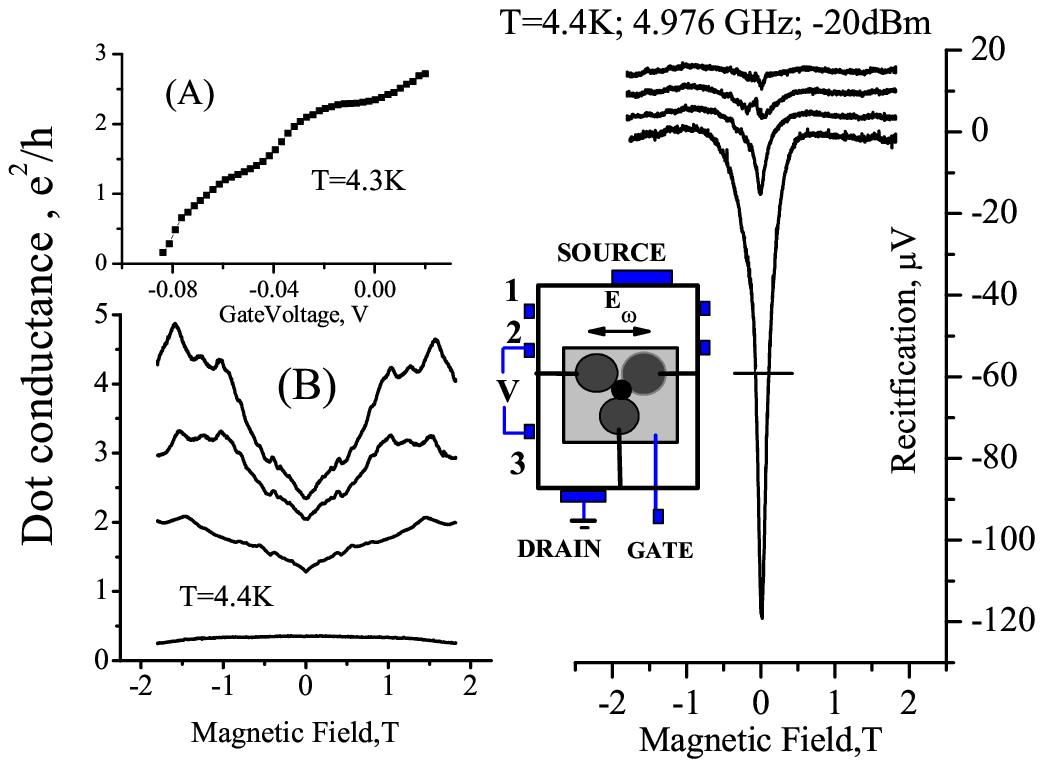} 
}
}
\vskip 0 cm
\hskip -0.2 cm
\small
\refstepcounter{figure}
\parbox[b]{3.3in}{\baselineskip=12pt FIG.~\thefigure.
Left panel: (A) Dependence of the dot conductance $G_{sd,23}$ on the gate voltage, H=0T. (B) Dependence of the dot conductance $G_{sd,23}$ on perpendicular magnetic field. Different curves correspond to different gate voltages
(from bottom to top in mV): -80.8, -52.9, -29.3, 0. Right panel: Dependence 
of the rectified voltage $V_{2-3}$ on  magnetic field at the same gate voltages 
as on the plot (B). Three upper curves are shifted up for clarity.  Sign of the rectified voltage corresponds to an electron pumping into the 2D area connected to contact 2 and the source.

\vspace{0.0in}
}
\label{1} 
\vskip 0.0cm
\normalsize
The increase of the conductance   is mainly due to the increase of electron transmission through the narrowest 0.2$\mu$, vertical conducting channel of the dot with the gate voltage increase.

 The rectification of the microwave radiation by the quantum dot is shown on the right panel of fig.1. The curves ($n=$2, 3 and 4) are shifted on 5$\times (n-1)$ $\mu$V up for clarity. Different curves correspond to different gate voltages. The ordering and the values of the voltages are the same as on the plot (B). The main and drastic property of the curves is the very strong dependence of the rectified voltage on the magnetic field. The reduction of the rectification by two order of magnitude is observed for the dot with the conductance below $e^2/h$. With an increase of the dot conductance the change of the rectified voltage with the magnetic field decreases. To estimate a scale of the magnetic field, which is important for the observed dependence, we measure the width $W$ of the first curve at half of it's height. This level is marked by a horizontal line in fig.1. The half width $W/2$=0.088 T is close to field H=0.085T at which the cyclotron radius  $r_H$ is equal to the dot size $d$ \cite{accuracy}. Thus the directed transport through the dot is reduced by half of it's zero field value at the magnetic field corresponding to condition $r_H \approx  d$.  The observation indicates that an electron ballistic motion inside the dot provides important contribution to the directed transport.

\vbox{
\vspace{0.3 in}
\hbox{
\hspace{-0.2in} 
\epsfxsize 3.3 in \epsfbox{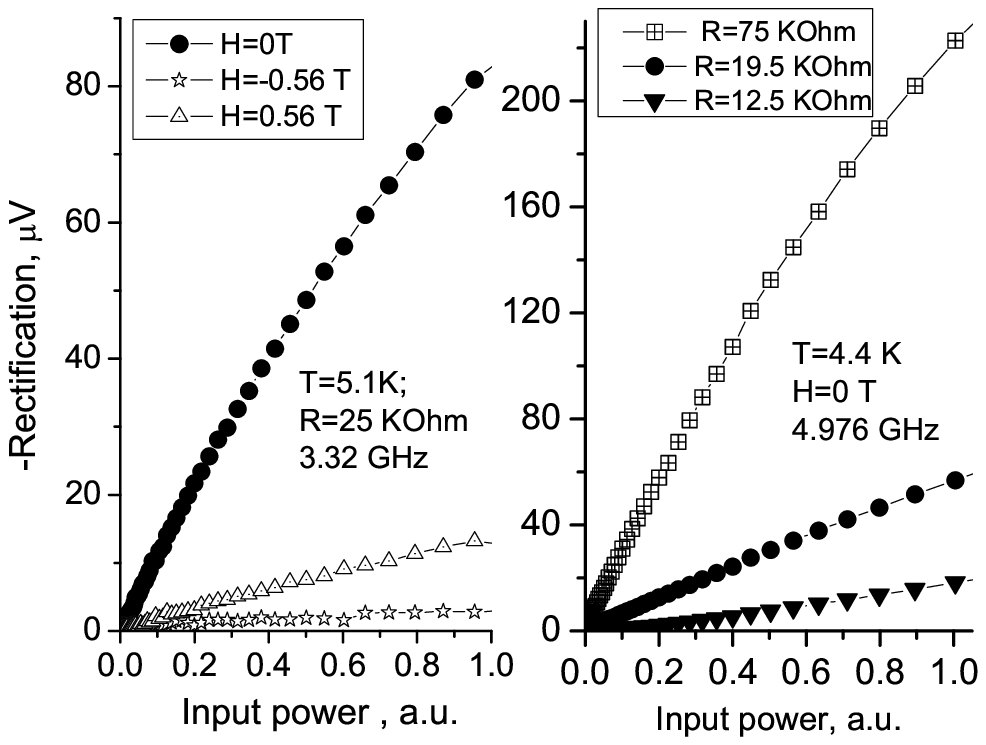} 
}
}
\vskip -0 cm
\hskip -0.2 cm
\small \refstepcounter{figure}
\parbox[b]{3.3in}{\baselineskip=12pt FIG.~\thefigure.
Left panel: Dependence of the rectification 
on microwave power 
at different magnetic field as labelled. 
The dot resistance $R_{sd,23}$ is 25 KOhm.
Right panel: dependence of the rectification on microwave power at 
different dot resistance at zero magnetic field.
The rectification sign is inversed.
\vspace{0.0in}
}
\label{2} 
\normalsize

A dependence of the rectified voltage on the microwave power is shown on figure 2 \cite{axes}. The left panel demonstrates the power dependence at different magnetic field H=-0.56 T, O T and +0.56 T for the dot with resistance $R_{sd,23}$=25 Kohm.
At zero magnetic field the rectified voltage below 50-60 $\mu $ V is proportional to the microwave power. A considerable sub-linear behavior is observed at higher level of the rectification. The right panel demonstrates the power dependence of the rectification at different resistance of the dot. A substantial reduction of the rectified voltage with the resistance is observed. It is partially related to the fact that the rectified voltage has to be proportional to the dot resistance: $V_{rect}=I_{rect} \times R$.

  Figure 3 demonstrates an evolution of the magnetic field dependence of the rectification with the microwave frequency. The symmetric (with respect to the magnetic field) part of the rectified voltage is presented on the left panel. A significant change of the form of the magnetic field dependence is observed at frequency above 10 GHz. An additional structure occurs at higher microwave frequencies, changing the direction of the rectified transport at zero magnetic field. The asymmetric part of the directed transport does not demonstrate any significant variations with the microwave frequency. The physical parameters relevant to the temporal behavior of the quantum dot are level spacing: $\Delta=2 \pi \hbar^2/(m^*A)=$28 mK, bulk momentum relaxation time: $\tau_p=$34 ps, dot crossing time $\tau_{cross}=(2-3)d/V_F=10-15$ (ps). The factor 2-3 is a result of the particular geometry of the dot. The dot crossing time appears to be the closest time at which the frequency dispersion of the directed transport begins to be observable:  $\omega \tau _{cross}\approx 1$ at $\omega/2\pi \approx$10 GHz. 
\vbox{
\vspace{0.2 in}
\hbox{
\hspace{-0.2in} 
\epsfxsize 3.3 in \epsfbox{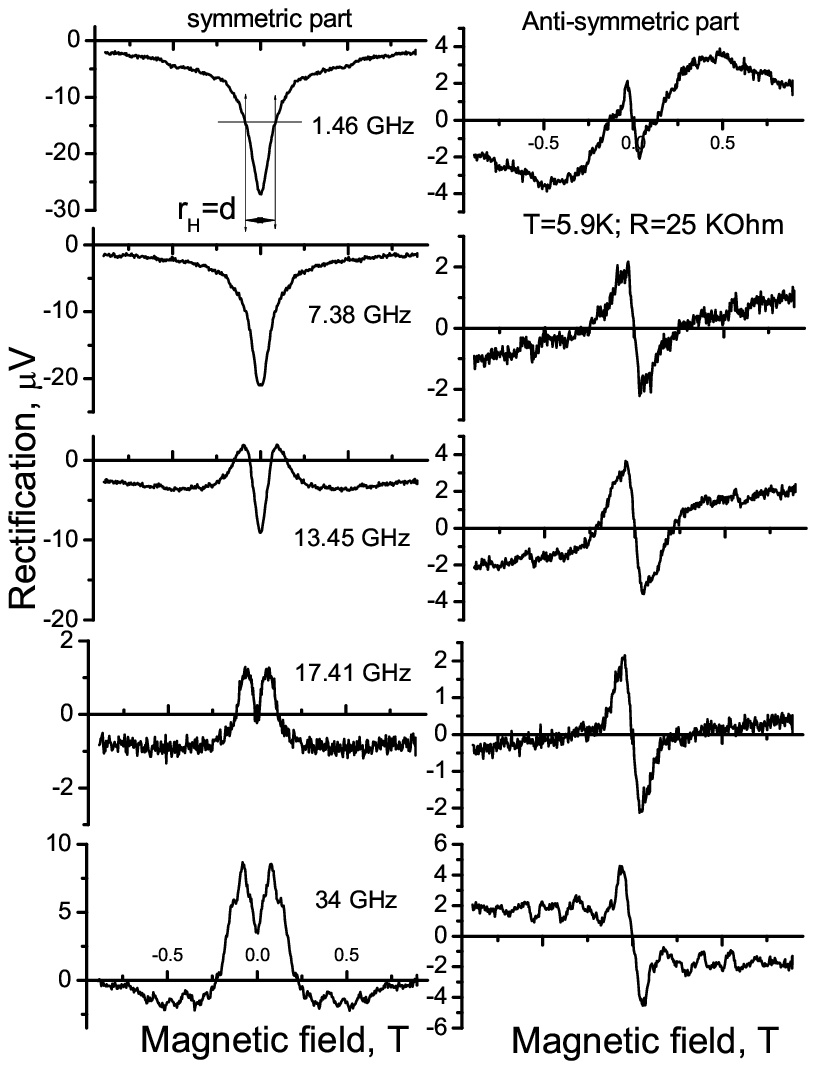} 
}
}
\vskip 1.5 cm
\hskip -0.2 cm
\small
\refstepcounter{figure}
\parbox[b]{3.3in}{\baselineskip=12pt FIG.~\thefigure.
Rectified voltage vs magnetic field at different microwave frequency as labeled. 
Different microwave power is applied at different frequency. Perturbative regime: 
$V_{rect}\sim$power.  Form of the curves does not depend on the power.
The left panel presents the symmetric part of the magnetic field dependence.
The right panel presents the asymmetric part. T=5.9K, R=25KOhm. 

\vspace{0.0in}
}
\label{3} 
\normalsize

At small applied power the rectified voltage   at zero magnetic field demonstrates rather weak temperature dependence. The voltage increases by 10-20 \%  with the temperature decrease from 6 K down to 0.3 K. The weak temperature dependence suggests again the ballistic nature of the observed phenomenon.  At low temperature (below 1 K) we have detected reproducible irregular variations of the dot conductance and the rectification with magnetic field and the gate voltage similar to reported earlier \cite{dicarlo}. This phenomenon is beyond the scope of the presented paper.   

Although the study of rectification in the mesoscopic quantum dot has already more than one decade history, the main efforts have been applied to investigate (both theoretically and experimentally) effects of quantum coherence on the nonlinear electron transport. Very little is studied ( at least experimentally ) on directed classical transport through  ballistic dots. However,
 in accordance with general principles \cite{belinicher} the directed transport through asymmetric constrictions may have considerable efficiency even in the classical case. In our experiment the main requirements for the directed transport: lack of a center of symmetry and an interaction with thermostat are fulfilled. 

Below we would like to discuss possible mechanisms, which may provide contributions to the directed transport through the dot at relatively high temperatures neglecting the quantum interference effects. One of the simplest mechanisms of the dot rectification is due to modulation of the gate voltage (and, therefore, the dot conductance, see fig.1 A ) by microwave electric field. In some sense the mechanism is a classical analog of the adiabatic charge transport. If the dot conductance is modulated in phase with the driving AC voltage $V_\omega$, then a net directed current is expected through the dot due to the simple fact, that the AC current during the first half of a voltage oscillation, when the dot conductance is high, is higher than the one during the second half of the oscillation (when the dot conductance is small). It is clear, that the net directed transport is proportional to the degree of the modulation of the dot conductance by the external source. 

We have checked the contribution of the discussed mechanism to the observed rectification. In a perturbative regime, at which the modulation of the dot conductance is relatively small, one can expect the net directed current to be proportional to the first derivative of the dot conductance to the gate voltage $V_g$: $I_{rect} \sim dG/dV_g$. We have measured the conductance of the dot as a function of the gate voltage $G(V_g)$ (see fig.1 A) and found it's first derivative. We compare the dependence of the derivative $dG(V_g)/dV_g$ on the gate voltage $V_g$ with the gate voltage dependence of the rectification signal measured in the perturbative regime. We have found correlations between $dG(V_g)/dV_g$ and the rectification as a function of the $V_g$ at very negative gates, at which the $dG(V_g)/dV_g$ is the strongest.  However, the amplitude of the correlated response was at least ten times less than the total rectification at T=4.4K. We have concluded that the mechanism, related to the gate modulation, although it is observed, does $not$ provide the main contribution to the dot rectification.    

The mechanism discussed above does not essentially use the ballistic electron motion. It may (and does) work even in completely disordered electron systems. Below we would like to propose a mechanism, which is efficient only for ballistic systems. Lets consider an equilateral triangular dot with three conducting channels placed at its corners: top, left and right \cite{simple}. Most of electrons ejected from the channels propagate through the dot scattering by the dot boundary. However there is a small fraction of electrons, which move   ballistically from one channel to another. Below we will name these electrons as effective electrons. The density of the effective electrons propagating from the left (right) channel to the top channel is $n_{l}, (n_{r})$.

The total electron flux due to the effective trajectories through the top channel $F_t$ is a sum of two fluxes from the left and right channels at the bottom of the triangle: $F_{t}=F_{l}+F_{r}$. At thermal equilibrium the flux $F_{t}$ is compensated completely by the electron flux $F_{b}$ from the  bulk 2D electron system connected to the external end of the top channel. Thus, at the equilibrium the net electron flux through the top channel is zero: $F_{net}=F_{t}-F_{b}=0$. Application of an AC electric field $E_\omega$ directed from left to right, parallel to the bottom of the triangle increases the velocity of the effective electrons from the left channel. To the first order in $E_\omega$ the increase is: $\delta v_{l}=v_{l}-v_F=\alpha E_\omega$, where $\alpha$ is a constant. Therefore, the flux $F_{l}$ increases by amount $\delta F_{l}=en_{l} \times \alpha E_\omega$. The electron flux from the right channel decreases by the same amount $\delta F_{r}=-en_{r} \times \alpha E_\omega$ due to the equilateral symmetry of the dot.  Therefore to the first order in  $E_\omega$ the flux $F_{t}$ remains to be the same as at the equilibrium.

To the second order in $E_\omega$, there is a nonzero contribution to the directed flux through the top channel. We believe, that this is the main reason of the net flux. The contribution is due to a $bending$ of the effective ballistic trajectories by the electric field $E_\omega$ inside the dot \cite{bending}. Due to the strong non-locality of the conductivity inside the dot, the direction of the electric field (determined by all electrons) does not, in general, coincide with the direction of the effective trajectories. The normal (to the trajectories) component of the electric field bends the ballistic trajectories. It changes the density of the effective electrons propagating from one channel to another (see below). A simple analysis demonstrates that the bending is opposite in the vertical direction for electrons coming from the left and right channels.  At small bending the displacement $\delta s$ of the end of an effective trajectory is proportional to the electric field $E_\omega$:
$\delta s \sim e E_\omega \tau_{cross}^2$. If the ballistic propagation of electrons from channels has a directivity (not isotropic), then the bending (displacement) of the ballistic trajectories should change the local density of the trajectories and, therefore, the density of the effective electrons. At small bending the change of the density should be proportional to the displacement of the trajectories and, therefore, to the electric field $E_\omega$. For the density of the effective electrons propagating from the left (right) channel to the top one we have found:  $n_{l}(n_{r})=n_0  \pm \beta E_{\omega}$, where the $\beta$ is a numerical constant.    Thus the net flux through the top channel is 

$$F_{t,net} =en_{l}\times \alpha E_\omega-en_{r}\times \alpha E_\omega- F_b= 2\alpha \beta E_\omega^2 \eqno{(1)}$$

In accordance with eq.1 the net flux is proportional to square of the electric field.  Thus the net current has non-zero value for the harmonically oscillating electric field with a zero mean.  

Important property of the discussed mechanism is a natural sensitivity of the directed transport to the magnetic field. A strong magnetic field affects substantially the ballistic trajectories, forcing the effective electrons to perform a cyclotron like, circular motion inside the dot. It should reduce considerably the flux transported by the effective trajectories. The magnetic field scale, at which the strong reduction should occur, corresponds to condition $2r_H < d$. At this condition an electron at Fermi level can perform circular motion inside the dot.  The scale is in a reasonable agreement with our experiment: $r_H \approx d$ (see fig.1 and fig.3).

In conclusion, an unexpected gigantic reduction of directed electron transport through a ballistic quantum dot under microwave radiation is observed. The magnetic field scale at which the reduction occurs satisfies condition $r_H \approx d$. It suggests strongly the ballistic origin of the observed phenomena. A ballistic model is proposed based on bending of electron trajectories inside the dot. There is no quantitative theory of this new phenomena.  

We thank J. Gertsen, D. Shepelyansky  and V. Fal'ko  for helpful discussions. The work is supported by NSF: DMR 0349049, by  RFBR : grant 02-05-16591,  NATO Linkage,  programs "Low-dimensional quantum structures" and "Quantum macrophysics" of RAS.

\end{multicols}


\begin{references} 
 
\bibitem{thouless} D. Thouless, Phys.Rev. B {\bf 27},6083 (1983).
\bibitem{falko} V. I. Falko and D. E. Khmel'nitskii, Zh.Eksp.Teor.Fiz. {\bf 95},328 (1989) [Sov. Phys.JETP {\bf 68}, 186 (1989)].
\bibitem{bykov} A.A.Bykov  {\it et al.}, Pis'ma Zh.Eksp.Teor.Fiz. {\bf 49}, 13, (1989)  [JETP Lett., {\bf 49}, 13 (1989)].
\bibitem{spivak} B. Spivak {\it et al.} Phys. Rev. B {\bf 51}, 13226 (1995)
\bibitem{brouwer} P.W. Brouwer, Phys.Rev. B {\bf 58}, R10135 (1998).
\bibitem{vavilov} M. G. Vavilov and I. L. Aleiner, Phys. Rev. B {\bf 64}, R16311 (1999). 
\bibitem{dicarlo} L. DiCarlo, C. M. Marcus and J. S. Harris, Jr., Phys. Rev. Lett, {\bf 91}, 246804 (2003). 
\bibitem{accuracy} The measurement of the width has been performed on a curve which is not symmetric with respect to H ( see fig.3). The functional dependence of the rectification on H is unknown. Therefore the agreement indicates just a  scale of the magnetic field relevant to the curve.   
\bibitem{axes} The x-axes represents the power applied to an input of the coaxial line. The actual power applied to sample is different (and frequency dependent). It has not been measured in the experiment. The functional dependence of the rectification on the power applied to the sample is the same as shown in the figure 2 due to linearity of the microwave circuit used in the experiment.      
\bibitem {belinicher} V. I Belinicher, and B. I. Sturman, Sov. Phys. Usp. {\bf 23}, 199 (1980).
\bibitem{simple} This apparent simplification of the actual dot geometry is used to present the main components of the proposed mechanism.      
\bibitem{bending} To the first order in $E_\omega$, the bending provides no net flux through the top channel. 

\end{references}
\end{document}